\documentstyle[12pt]{article}

\newlength{\dinwidth}
\newlength{\dinmargin}
\begin{document}
%~~\\
%%\vspace{1cm}
\begin{center}
\begin{Large}
\begin{bf}
%
% TITLE
%
Measuring~~$|V_{ub}|$~~at~~the~~LEP/SLC
\footnote{Talk given by C.S. Kim, at the
Int. Workshop on B-physics, Nagoya, Japan, on Oct. 26--28, 1994.
Proceedings published by World Scientific, Singapore, edited by A. Sanda and
S. Suzuki.}
\\
\end{bf}
\end{Large}
\vspace{5mm}
\begin{large}
%
% Your NAME
%
%Author NAME
C.S.~Kim$^{ a}$~~ and ~~ A.D.~Martin$^{ b}$
\\
\end{large}
$a$: Department of Physics, Yonsei University, Seoul 120--749, Korea\\
$b$: Department of Physics, Durham University, Durham DH1 3LE, England\\
\vspace{5mm}
\end{center}
%\begin{quotation}
\noindent
%
% ABSTRUCT
%
\begin{abstract}

We show that the observation of energetic charged leptons from
semileptonic $B$-meson decays at the LEP and SLC $Z$--boson factories offers
a unique opportunity to measure the quark mixing matrix element
$|V_{ub}|$.  We present various distributions of the $B$ decay products
to show that the $b \to u$ decays can be cleanly separated from
the $b \to c$ decays, with a signal of $O(100)$ events per million
$Z$ bosons produced.

\end{abstract}

Our knowledge of the CKM or quark mixing matrix elements $V_{cb}$ and
$V_{ub}$ comes from studies of the semileptonic decays of $B$ mesons.
The larger element, $V_{cb}$, is well detemined both from inclusive
$B \to X_c l \nu$ decays and from the exclusive $B \to D^* l \nu$ decay.
The measurements have been obtained from studies of $B \bar B$ production
in $e^+ e^-$ collisions at $\Upsilon (4S)$~\cite{1,2} and at the $Z$ boson
resonance~\cite{3}.
The results have been reviewed recently by Patterson~\cite{4}.
Using the inclusive decays it is found
\begin{eqnarray}
|V_{cb}| &=& 0.039 \pm 0.001 \pm 0.005~~~~
[~\mbox{at}~\Upsilon (4S)~]\ \nonumber\\
|V_{cb}| &=& 0.042 \pm 0.002 \pm 0.005~~~~
[~\mbox{at}~Z~]\ \nonumber
\end{eqnarray}
where the first error is experimental and
the second is due to the dependence
on theoretical models. The observation of the exclusive $B \to D^* l \nu$
decay gives
\begin{displaymath}
|V_{cb}| = 0.040 \pm 0.002 \pm 0.002
\end{displaymath}
where here the theoretical error is smaller since the form factors for
the $b \to c$ (`heavy-quark $\to$ heavy-quark') transition are well studied
and under control~\cite{5}.  Model dependence only enters at the level of
power corrections, which are suppressed by at least a factor of
$(\Lambda _{QCD}/m_c)^2$, see Ref.~\cite{6} for a recent review.

The situation for $V_{ub}$ is very different. The present information
comes from $\Upsilon (4S) \to B \bar B$. The extraction of
$V_{ub}$ is based on the fact that only $b \to u l \nu$ is kinematically
able to populate the high end of the inclusive momentum spectrum of the
charged lepton $l$. The much more numerous $b \to c l \nu$ decays have
$p(l) < 2.5$ GeV. By determining the small excess of events over the off
resonance continuum in the short interval $2.5 < p(l) < 3$ GeV it is found
that~\cite{4}
\begin{displaymath}
\left|{V_{ub} \over V_{cb}}\right| = 0.09 \pm 0.03
\end{displaymath}
where the error is almost entirely due to the sensitivity of the
analysis to the various theoretical models. Recently it has been proposed
that measurements of the hadronic invariant mass~\cite{7,8} or hadronic
energy spectrum~\cite{9} of inclusive semileptonic $B$ decays may be useful
in extracting $|V_{ub}|$ in a less model dependent manner. In future
asymmetric $B$--factories, with vertex detection, these offer alternative
ways of identifying $b \to u$ transions.

Turning now to the exclusive decays, we note that the rates for
$B \to \rho l \nu,~ \omega l \nu$ are so low that only the upper limit
{}~\cite{1,2,4}
\begin{displaymath}
\left|{V_{ub} \over V_{cb}}\right| < 0.10 - 0.13
\end{displaymath}
has been obtained, where the range indicates the model dependence.
Even if appreciable statistics for the exclusive channels become available,
the theoretical description will be more problematic than $b \to c$
since $b \to u$ is a `heavy-quark $\to$ light-quark' transition.

In this letter we investigate the possibility of measuring $|V_{ub}|$
at $Z$--factories (LEP, SLC) by searching for particular distinctive
kinematic characteristics of inclusive $B \to X_u l \nu$ decays. We have
\begin{displaymath}
\Gamma(Z \to b \bar b) \simeq 0.38~ \mbox{GeV}
\end{displaymath}
with a branching fraction of about $15 \%$ of $Z$ decays leading to
$b \bar b$ final states~\cite{10}. For the purposes of
illustration we will assume that the
`$b$-quark $\to$ $B$-meson' fragmentation is described by the Peterson
{\it et. al.} function~\cite{11}
\begin{displaymath}
f(x) = N {x (1-x)^2 \over \left[(1-x)^2 + \epsilon_p x \right]^2}
\end{displaymath}
with $x=|{\bf p}(B) / {\bf p}(b)|$. We allow the parameter $\epsilon_p$
to span the range $\epsilon_p = 0.006 \pm 0.003$.
We use the ACCMM model~\cite{12} to relate the inclusive $B \to X_q l \nu$
decays (with $q=u$ or $c$) to the $b \to q l \nu$ transition. In this
model the Fermi motion of the $b$ quark inside $B$ meson is assumed
to correspond to a Gaussian momentum distribution
\begin{displaymath}
\phi({\bf p}) = {4 \over \sqrt{\pi} p_{_F}^3}
\mbox{exp}(-{\bf p}^2/p_{_F}^2)
\end{displaymath}
where $p_{_F} \simeq 0.3$ GeV is widely used in analyses. To investigate
the sensitivity to $p_{_F}$ we also take  $p_{_F} = 0.5$ GeV, a value
found in a relativistic quark model calculation~\cite{13}.
In the ACCMM model we have
\begin{displaymath}
d \Gamma_B (p_{_F},m_{sp},m_q) = \int_0^{p_{max}} dp~p^2 \phi({\bf p})
d \Gamma_b (m_b)
\end{displaymath}
where $p_{max}$ is the maximum kinetically allowed value of
$p = |{\bf p}|$. We take $m_c = 1.6$ GeV for $b \to c$ decay and
$m_u = 0.15$ GeV for $b \to u$, and we ensure the correct hadronic
end-point by assuming that the spectator quark mass to be
$m_{sp} = m_{_D} - m_c$ for $b \to c$ and $m_{sp} = m_{\pi} - m_u
\approx 0$ GeV for $b \to u$.  To investigate the sensitivity of the
results to the choice of $m_u$ we repeated the
calculation with $m_u = 0.3$ GeV (keeping
$m_{sp} = 0$).

The results are given in Figs. 1--2, where for each kinematic variable
($x \equiv E(l),~E(X)$ and $p_{_T}^{^X}(l)$) we show the normalized
distribution
\begin{displaymath}
{1 \over \Gamma(Z \to b \bar b)}~{d \Gamma \over dx}
\left(Z \to b \bar b~[~\to (B \to X_q l \nu)~]~\right)
\end{displaymath}
in units of BR$(B \to X_q l \nu)/$GeV with $q=c$ or $u$ and
$l=e^-$. In other words the total integral over each distribution is
normalized to unity when no cuts are applied (and GeV units are used).
To the right-hand-side of each figure we show the event numbers/GeV
for $10^6~Z$ bosons produced. We take
BR$(B \to X_c e^- \bar \nu) = 10.4 \%$ [10] and we assume
$|V_{ub}/V_{cb}| = 0.1$. If we include the $e^+$ and $\mu^\pm$ final
states then the rate shown is multiplied by a factor of 4.
The variables that are most relavant for the extraction of $|V_{ub}|$ from
$B \to X l \nu$ decays are the charged lepton energy $E(l)$, the
energy of the hadronic system $E(X)$ and the perpendicular momentum
$p_{_T}^{^X}(l)$ of the charged lepton with respect to the hadronic jet $X$.
We use the $Z$ rest frame.

Fig. 1 shows the energy spectrum of the charged lepton, $E(l)$. In Fig. 1(a)
no cuts are applied and we see that events with $E(l) \geq
35$ GeV originate dominantly from $b \to u$ transitions. Indeed for
$E(l) > 35$ GeV we have a very clean sample
$B \to X_u l \nu$ decays corresponding to
$\sim 40$ $b \to u$ events per $10^6~Z$ bosons produced,
taking $l=e^\pm,\mu^\pm$.

We can increase the $b \to u$ sample by observing the hadronic jet $X$.
In Fig. 1(b) we again show the charged lepton spectrum
but now with the hadronic energy cut $E(X) > 5$ GeV and the perpendicular
momentum cut $p_{_T}^{^X}(l) > 5$ GeV imposed. Indeed these hadronic cuts
have the effect of eliminating $b \to c$ events completely.
Even if we conservatively consider only events with $E(l) > 25$ GeV
in order to eliminate any cascade
events from $b \to c \to s l \nu$, then Fig. 1(b) shows we should
have a clean $b \to u$ sample of $\sim 110$ events per $10^6~Z$ bosons
produced.  With these leptonic and hadronic cuts ($E(l) > 25$ GeV and
$E(X), p_{_T}^{^X}(l) > 5$ GeV)  the momentum flow to the unobservable
neutrino  is kinematically restricted,
so we can use
the direction of the other side $B$ (or $\bar B$) to determine
$p_{_T}^{^X}(l)$ more correctly even with the loss of neutrals.

The effects of the hadronic cuts are shown in Figs. 2(a), (b). Fig. 2(a)
illustrates
the discriminatory power of the $p_{_T}^{^X}(l)$ distribution in selecting
$b \to u$ events. However, this variable can only be employed if the
hadronic jet $X$ is identified, as explained above, and we show
in Fig. 2(a) the effect of imposing cuts of $E(l) > 25$ GeV and
$E(X) > 5$ GeV. For completeness
we show in Fig. 2(b) the $E(X)$ spectrum with $p_{_T}^{^X}(l) > 5$ GeV and
$E(l) > 25$ GeV.

A few remarks are in order. (a) Although the invariant mass
of the decay hadrons is the best observable~\cite{7}
to separate the
$b \to u$ from the $b \to c$ events,
its precise measurement {\it at $Z$ factories} with the large hadronic
environment is difficult
%(even with a vertex detector to identify the decay hadrons)
since it requires knowledge of all the momentum components of the decay
hadrons,
and there could well be neutral particles which escape detection.
(b) We explored
the possible model dependence by varying the input parameters in the ranges:
$\epsilon_p=0.006 \pm 0.003$, $m_u=0.15 - 0.3$ GeV and $p_{_F}=0.3 - 0.5$
GeV. We also adopted normalized distributions to eliminate theoretical
uncertainty from total event rates of $Z \to b \bar b$.
We found very little difference in the shape of the spectrum of
the $b \to u$ decays, as shown in figures.
For the $b \to c$ transition, we can see a pronounced difference in shape
as we vary the input parameters over the ranges.
(c) Recently, Bigi {\it et. al.}~\cite{14} proposed a model
independent decay spectrum using heavy quark effective theory by including
nonperturbative $1 / m_{Q}^2$ corrections. They found
that their spectrum can be
well reproduced by the ACCMM model with $p_{_F} \simeq 0.3$ GeV.
This model~\cite{14} cannot describe properly the end-point region,
$\delta E(l) \approx 300$ MeV, of $E(l)$ spectrum, which is the
important region for measuring $|V_{ub}|$ at the $\Upsilon (4S)$.
And this has been a subject of
recent investigations. It has been shown by Ref.~\cite{15} that the end-point
region is described by a universal light cone distribution function,
and that the ACCMM model may be rewritten in terms of such a distribution
function.
At the $Z$ resonance, since we can use at least
events with $35 \leq E(l) \leq 45$ GeV, we can
avoid any difficulty in applying this model.
(d) There could also be
some model dependence from `$b$-quark $\to$ $B$-meson' fragmentation.
With the large $b$ production rate at the $Z$ resonance, this is the best place
to study the fragmentation effect of the heavy quark.

In conclusion we note that the observation of very energetic charged
leptons ($E(l) > 35$ GeV) in $b \bar b$ production at $Z$ factories offers
a sizeable clean sample of $b \to u$ events, with little or no
contamination from $b \to c$ processes. Moreover statistics can be
significantly increased by the observation and measurement of the hadronic
system accompanying the decay. Particular event topologies are shown
which eliminate the $b \to c$ transitions regardless, in principle, of
the lepton energy $E(l)$. However, in practice, it will be
best to use Monte Carlo studies to optimize both
the hadronic and leptonic cuts so as to extract the cleanest possible
$b \to u$ data sample.

\vspace{0.8cm}
\noindent
This work  was supported
in part by the Korean Science and Engineering  Foundation,
in part by Non-Direct-Research-Fund, Korea Research Foundation 1993,
in part by the Center for Theoretical Physics, Seoul National University,
in part by Yonsei University Faculty Research Grant, and
in part by the Basic Science Research Institute Program, Ministry of
Education, 1994,  Project No. BSRI-94-2425.

%
% REFERENCES
%

\vfill\eject
%\end{document}

\vspace*{15.5cm}
\noindent
Fig.1 (a) The normalized energy distribution of electron,
$1/\Gamma(Z \to b \bar b) d\Gamma(Z \to b \bar b[\to
(B \to X_q e^- \bar \nu)])/ d E(e^-)$ in units of
BR$(B \to X_q e^- \bar \nu)/$GeV with $q=c$ or $u$.
On the right-hand-side of the figure we show the event numbers/GeV
for $10^6~Z$ bosons produced. We take BR$(B \to X_c e^- \bar \nu) =
10.4 \%$ [10] and we assume that
$|V_{ub}/V_{cb}| = 0.1$. If we include the $e^+$ and $\mu^\pm$ final
states then the rate shown is multiplied by a factor of 4.
The narrow band, visible only for $B \to X_c$, results from
varying the input parameters over the ranges
$\epsilon_p=0.006 \pm 0.003$, $m_u=0.15 - 0.3$ GeV and
$p_{_F}=0.3 - 0.5$ GeV.  We take $m_c=1.6$ GeV.
(b) The normalized $E(e^-)$ distribution of $B \to X_u e^- \bar \nu$
after the imposition of the cut, $E(X) > 5$ GeV, on the energy of
hadronic system and the cut, $p_{_T}^{^X}(e^-) > 5$ GeV,
on the perpendicular momentum of electron with respect to the hadronic jet.
\vfill\eject

\vspace*{19.5cm}
\noindent
Fig.2  (a) The normalized $p_{_T}^{^X}(e^-)$ distribution with
cuts $E(e^-) > 25$ GeV and $E(X) > 5$ GeV imposed.
The narrow band arises as in Fig. 1.
(b) The normalized $E(X)$ distribution with cuts
$E(e^-) > 25$ GeV and $p_{_T}^{^X}(e^-) > 5$ GeV imposed.
\vfill\eject

\end{document}